\documentclass[a4paper,12pt,draftcls,peerreview,onecolumn]{IEEEtran}%
\usepackage{amsfonts}
\usepackage{amsmath}
\usepackage{amssymb}
\usepackage{graphicx}%
\providecommand{\U}[1]{\protect\rule{.1in}{.1in}}
\begin{document}

\title{A New Causal Ideal Internal Dynamics Generator}
\author{Quan Quan and Kai-Yuan Cai\thanks{Corresponding Author: Quan Quan, Associate
Professor, Department of Automatic Control, Beijing University of Aeronautics
and Astronautics, Beijing 100191, qq\_buaa@buaa.edu.cn,
http://quanquan.buaa.edu.cn.}}
\maketitle

\begin{abstract}
The design of ideal internal dynamics (IID) generators, namely solving IID, is
a fundamental problem, which is a key step to handle the nonminimum-phase
output tracking problem. In this paper, for a class of unstable matrix
differential equations, a new causal dynamic IID generator is proposed, whose
parameters are partly chosen via $\mathcal{H}_{2}/\mathcal{H}_{\infty}$
optimization. Compared with existing similar generators, it is applicable to
matrix differential equations with singular system matrices and is easily
extended to slowly time-varying matrix differential equations without extra computation.

\end{abstract}

\begin{keywords}
Nonminimum-phase systems, ideal internal dynamics, causal case, tracking.
\end{keywords}

\section{Introduction}

A system is nonminimum-phase if its internal dynamics (ID) are unstable
\cite{Isidori(1995)}. Nonminimum-phase output tracking is a challenging,
real-life control problem that has been extensively studied. An important way
for this problem is to identify the state references such that the output
tracking problem can be converted to be an easier stabilization problem, which
can be solved by using conventional control methods, such as sliding mode
control methods \cite{Gopalswamy(1993)},\cite{Shkolnikov(2002)}. State
references are composed of output references and internal state references.
The former are often given, whereas the latter is difficult to obtain for an
unstable ID, namely for a nonminimum-phase system. A bounded solution to the
unstable ID is called the ideal internal dynamics (IID)
\cite{Gopalswamy(1993)}. A\emph{ }basic\emph{ IID Problem} can be stated as:

\emph{IID Problem:} \emph{Given }$\xi\in\mathcal{L}_{\infty}\left(  \left[
0,\infty\right)  ,%
%TCIMACRO{\U{211d} }%
%BeginExpansion
\mathbb{R}
%EndExpansion
\right)  $\footnote{$f\in\mathcal{L}_{\infty}\left(  \left[  0,\infty\right)
,%
%TCIMACRO{\U{211d} }%
%BeginExpansion
\mathbb{R}
%EndExpansion
^{n}\right)  $ denotes that $f\left(  t\right)  \in%
%TCIMACRO{\U{211d} }%
%BeginExpansion
\mathbb{R}
%EndExpansion
^{n}$ and sup$_{t\geq0}\left\Vert f\left(  s\right)  \right\Vert <\infty.$%
},\emph{ }$A\in%
%TCIMACRO{\U{211d} }%
%BeginExpansion
\mathbb{R}
%EndExpansion
^{n\times n}$\emph{ and }$N\in%
%TCIMACRO{\U{211d} }%
%BeginExpansion
\mathbb{R}
%EndExpansion
^{n}$,\emph{ find an initial condition }$\eta_{0}$\emph{ such that the
solution }$\eta\left(  t\right)  $\emph{ to the following differential
equation}%
\begin{equation}
\dot{\eta}\left(  t\right)  =A\eta\left(  t\right)  +N\xi\left(  t\right)
,\eta\left(  0\right)  =\eta_{0},t\geq0 \label{easy}%
\end{equation}
\emph{belongs to} $\mathcal{L}_{\infty}\left(  \left[  0,\infty\right)  ,%
%TCIMACRO{\U{211d} }%
%BeginExpansion
\mathbb{R}
%EndExpansion
\right)  $\emph{.}

The \emph{IID Problem }is in fact about the noncausal (offline) case, where
$\xi\left(  s\right)  ,$ $s\in\left[  0,\infty\right)  $ is available before
finding\emph{ }the\emph{ }solution\emph{ }$\eta.$ If $A$ is stable, then the
IID can be obtained by solving the differential equation (\ref{easy}) directly
in forward time, whereas it cannot for an unstable $A$. For an unstable $A$,
the basic idea of solving the IID with an unstable $A$ is to run the stable
parts forward in time and the unstable parts backward with the priori
information. However, it does not work in the the causal (online) case, where
only $\xi\left(  s\right)  ,$ $s\in\left[  0,t\right]  $ is available to
determine\emph{ }the\emph{ }solution\emph{ }$\eta$ at the time $t$. This
problem can be formulated in general as:

\emph{Causal IID Problem: Given }$\xi\in\mathcal{L}_{\infty}\left(  \left[
0,T\right]  ,%
%TCIMACRO{\U{211d} }%
%BeginExpansion
\mathbb{R}
%EndExpansion
\right)  $\emph{, }$\hat{\eta}_{T}\left(  0\right)  =0,$ $A\in%
%TCIMACRO{\U{211d} }%
%BeginExpansion
\mathbb{R}
%EndExpansion
^{n\times n}$\emph{, }$N\in%
%TCIMACRO{\U{211d} }%
%BeginExpansion
\mathbb{R}
%EndExpansion
^{n}$ \emph{and }$\delta>0,$\emph{ find a differentiable function }$\hat{\eta
}_{T}\in\mathcal{L}_{\infty}\left(  \left[  0,T\right]  ,%
%TCIMACRO{\U{211d} }%
%BeginExpansion
\mathbb{R}
%EndExpansion
^{n}\right)  $\emph{ such that}\footnote{$\mathcal{B}\left(  \delta\right)
\triangleq\left\{  \xi\in%
%TCIMACRO{\U{211d} }%
%BeginExpansion
\mathbb{R}
%EndExpansion
\left\vert \left\Vert \xi\right\Vert \leq\delta\right.  \right\}  ,$
$\delta>0;$ the notation $x\left(  t\right)  \rightarrow\mathcal{B}\left(
\delta\right)  $ means $\underset{y\in\mathcal{B}\left(  \delta\right)  }%
{\inf}$ $\left\vert x\left(  t\right)  -y\right\vert \rightarrow0.$}%
\[
\dot{\hat{\eta}}_{T}\left(  T\right)  -A\hat{\eta}_{T}\left(  T\right)
-N\xi\left(  T\right)  \rightarrow\mathcal{B}\left(  \delta\right)  \text{, as
}T\rightarrow\infty.
\]

In \cite{Gopalswamy(1993)}, the noncausal IID problem was considered for a
class of forcing terms generated by a known nonlinear exosystem. The problem
was further solved for a class of more general systems and a class of more
general forcing terms in \cite{Devasia(1996)}. However, these inversion-based
approaches require the entire output references ahead of time which restricts
the use. To overcome this limitation, the preview-based stable-inversion
approaches were proposed \cite{Zou(1999)},\cite{Zou(2009)}. It requires the
finite-previewed (in time) future output reference and thus enables the online
implementation. Such a problem can be formulated as a modified\emph{ Causal
IID Problem }that\emph{ }finds a solution $\hat{\eta}_{T}\in\mathcal{L}%
_{\infty}\left(  \left[  0,T\right]  ,%
%TCIMACRO{\U{211d} }%
%BeginExpansion
\mathbb{R}
%EndExpansion
^{n}\right)  $ by $\xi\in\mathcal{L}_{\infty}\left(  \left[  0,T+T_{pre}%
\right]  ,%
%TCIMACRO{\U{211d} }%
%BeginExpansion
\mathbb{R}
%EndExpansion
\right)  $, where $T_{pre}>0$ is the preview time. It has been shown that a
large enough preview time is critical to ensure the precision in the
preview-based output tracking. However, for some cases, the forcing term
$\xi\left(  t\right)  $ in (\ref{easy}) may be an online estimate of
uncertainties, namely the future information is unavailable. Therefore, the
solution idea for the noncausal IID is inapplicable to the causal IID problem.
To the best of our knowledge, the solutions to the causal IID problem are only
limited to a class of bounded forcing term generated by an exosystem. For a
class of forcing term generated by a linear exosystem, the IID can be given
exactly by solving a Sylvester equation proposed in \cite{Francis(1976)}. For
a nonlinear exosystem, we have to resort to a first-order partial differential
equation proposed in \cite{Isidori(1990)}. The two resulting IID generators
can generate the IID directly, which can be considered as static IID
generators. However, they require full knowledge of the state of the
exosystem, which however may not be obtained directly. Moreover, the resulting
IID will preserve the noise if the state of the exosystem is noisy. For these
reasons, the authors suppose, a dynamic IID generator was proposed to solve
the IID for the equation (\ref{easy}) in \cite{Shkolnikov(2002)}. Furthermore,
by using higher-order sliding mode differentiators, it was modified in
\cite{Shtessel(2010)} for an unknown matrix $A.$ However, both dynamic
generators do not cover the case that $A$ is singular as they require
obtaining $A^{-1}$. Furthermore, in the case of a time-varying matrix, they
will be time-consuming. For example, if adopt $\frac{d}{dt}A^{-1}\left(
t\right)  =-A^{-1}\left(  t\right)  \frac{d}{dt}A\left(  t\right)
A^{-1}\left(  t\right)  $\footnote{$A\left(  t\right)  A^{-1}\left(  t\right)
=I_{n}\Rightarrow\frac{d}{dt}A\left(  t\right)  A^{-1}\left(  t\right)
+A\left(  t\right)  \frac{d}{dt}A^{-1}\left(  t\right)  =0_{n\times
n}\Rightarrow\frac{d}{dt}A^{-1}\left(  t\right)  =-A^{-1}\left(  t\right)
\frac{d}{dt}A\left(  t\right)  A^{-1}\left(  t\right)  $} to generate
$A^{-1}\left(  t\right)  $ online, then we have to calculate about $n^{2}$
differential equations. The same difficulty also exists in solving a
time-varying Sylvester equation.

In this paper, we propose a new causal dynamic IID generator for a class of
perturbed forcing terms generated by linear exosystems. Analysis shows that
the equation (\ref{easy}) is solvable if $A$ is singular under the conditions
consistent with that for\ the Sylvester equation proposed in
\cite{Francis(1976)}. Furthermore, to suppress the perturbation by the noise,
the parameters are partly chosen via $\mathcal{H}_{2}/\mathcal{H}_{\infty}$
optimization so that the error bound caused by the perturbation can be
evaluated. To show the advantage, the proposed IID generator is also applied
to a slowly time-varying unstable differential equation in the simulation.
Compared with existing similar generators, it avoids computing $A^{-1}$ so
that it can cover the case that $A$ is singular, and is further easier to
apply to matrix differential equations with slowly time-varying system
matrices. Moreover, the proposed dynamic IID generator only needs to calculate
about $n$ differential equations. This reduces the computational complexity.
Finally, it should be pointed out that the proposed IID generator can also be
applied to the tracking problem for nonlinear nonminimum-phase systems by
following the idea as in \cite{Shkolnikov(2002)},\cite{Shtessel(2010)}, i.e.,
to lump weakly nonlinear terms and uncertainties into the forcing term $\xi.$

\section{Problem Formulation and Preliminary Results}

\subsection{Problem Formulation}

Consider the following unstable matrix differential equation:%
\begin{equation}
\dot{\eta}=A\eta+N\xi,\eta\left(  0\right)  =0 \label{unequation}%
\end{equation}
where

$-$ $\eta\in%
%TCIMACRO{\U{211d} }%
%BeginExpansion
\mathbb{R}
%EndExpansion
^{n}$ is the state;

$-$ $\xi\in\mathcal{L}_{\infty}\left(  \left[  0,\infty\right)  ,%
%TCIMACRO{\U{211d} }%
%BeginExpansion
\mathbb{R}
%EndExpansion
\right)  $ (it will be extended to be a vector later) could be modeled as
follows:%
\begin{equation}
\dot{w}=Sw,\xi=E^{T}w \label{exosystem}%
\end{equation}
where $w\in%
%TCIMACRO{\U{211d} }%
%BeginExpansion
\mathbb{R}
%EndExpansion
^{m},S\in%
%TCIMACRO{\U{211d} }%
%BeginExpansion
\mathbb{R}
%EndExpansion
^{m\times m},E\in%
%TCIMACRO{\U{211d} }%
%BeginExpansion
\mathbb{R}
%EndExpansion
^{m}$; here we consider the causal case, namely the signal $\xi\left(
s\right)  ,$ $s\in\left[  0,t\right]  $ is available at the time $t>0.$

$-$ $N\in%
%TCIMACRO{\U{211d} }%
%BeginExpansion
\mathbb{R}
%EndExpansion
^{n}$, and $A\in%
%TCIMACRO{\U{211d} }%
%BeginExpansion
\mathbb{R}
%EndExpansion
^{n\times n}$ is a non-Hurwitz matrix.

Denote $\hat{\eta}$ to be the estimate. The objective is to obtain a bounded
estimate $\hat{\eta}$ such that $y\left(  t\right)  =\dot{\hat{\eta}}\left(
t\right)  -A\hat{\eta}\left(  t\right)  -N\xi\left(  t\right)  \rightarrow0$
as $t\rightarrow\infty.$ Furthermore, consider the case that $\xi$ is a vector.

Before proceeding further with the development of this work, the following
preliminary result is needed.

\textit{Lemma 1}. If and only if rank$\left(  F-\lambda I_{n}\right)  =n-1$
for every eigenvalue $\lambda\in%
%TCIMACRO{\U{2102} }%
%BeginExpansion
\mathbb{C}
%EndExpansion
$ of $F\in%
%TCIMACRO{\U{211d} }%
%BeginExpansion
\mathbb{R}
%EndExpansion
^{n\times n},$ then there exists a vector $B\in%
%TCIMACRO{\U{211d} }%
%BeginExpansion
\mathbb{R}
%EndExpansion
^{n}$ such that the pair $\left(  F,B\right)  $ is controllable\footnote{It
should be noted that this useful property was first shown by Wonham
\cite{Wonham(1967)}. Later, the proof was simplified by Antsaklis
\cite{Antsaklis(1978)} in a completely different way. We have completed this
proof based on some basic knowledge on matrix before knew these previous
proofs. So, our proof is completely different from those in
\cite{Wonham(1967)},\cite{Antsaklis(1978)}.}.

\textit{Proof. See Appendix A.}

\section{A New Causal Ideal Internal Dynamics Generator}

Our IID generator is proposed as follows:
\begin{subequations}
\label{Augment}%
\begin{align}
\dot{x}  &  =A_{cl}x+N_{cl}\xi,x\left(  0\right)  =0\label{A1}\\
\hat{\eta}  &  =C_{cl}^{T}x \label{A2}%
\end{align}
where%
\end{subequations}
\begin{align*}
x  &  =\left[
\begin{array}
[c]{c}%
v\\
\hat{\eta}\\
e
\end{array}
\right]  \in%
%TCIMACRO{\U{211d} }%
%BeginExpansion
\mathbb{R}
%EndExpansion
^{m+n+1},v\in%
%TCIMACRO{\U{211d} }%
%BeginExpansion
\mathbb{R}
%EndExpansion
^{m},\hat{\eta}\in%
%TCIMACRO{\U{211d} }%
%BeginExpansion
\mathbb{R}
%EndExpansion
^{n},e\in%
%TCIMACRO{\U{211d} }%
%BeginExpansion
\mathbb{R}
%EndExpansion
,C_{cl}=\left[
\begin{array}
[c]{c}%
0_{m\times n}\\
I_{n}\\
0_{1\times n}%
\end{array}
\right]  \in%
%TCIMACRO{\U{211d} }%
%BeginExpansion
\mathbb{R}
%EndExpansion
^{\left(  m+n+1\right)  \times n},\\
A_{cl}  &  =\left[
\begin{array}
[c]{ccc}%
S & 0_{m\times n} & L_{11}\\
0_{n\times m} & A & L_{12}\\
L_{21} & L_{22} & L_{3}%
\end{array}
\right]  \in%
%TCIMACRO{\U{211d} }%
%BeginExpansion
\mathbb{R}
%EndExpansion
^{\left(  m+n+1\right)  \times\left(  m+n+1\right)  },N_{cl}=\left[
\begin{array}
[c]{c}%
0_{m\times1}\\
N\\
0
\end{array}
\right]  \in%
%TCIMACRO{\U{211d} }%
%BeginExpansion
\mathbb{R}
%EndExpansion
^{m+n+1},\\
L_{11}  &  \in%
%TCIMACRO{\U{211d} }%
%BeginExpansion
\mathbb{R}
%EndExpansion
^{m},L_{12}\in%
%TCIMACRO{\U{211d} }%
%BeginExpansion
\mathbb{R}
%EndExpansion
^{n},L_{21}\in%
%TCIMACRO{\U{211d} }%
%BeginExpansion
\mathbb{R}
%EndExpansion
^{1\times m},L_{22}\in%
%TCIMACRO{\U{211d} }%
%BeginExpansion
\mathbb{R}
%EndExpansion
^{1\times n},L_{3}\in%
%TCIMACRO{\U{211d} }%
%BeginExpansion
\mathbb{R}
%EndExpansion
.
\end{align*}
The basic idea is to make (\ref{Augment})\ satisfy the following two conditions:

i) $A_{cl}$ is stable;

ii) $e\left(  t\right)  \rightarrow0$ as $t\rightarrow\infty.$

By taking $\xi$ as the input and $x$ as the state, the condition i) implies
the bounded-input bounded-state stability of (\ref{A1}), namely the resulting
$\hat{\eta}$ is bounded. On the other hand, (\ref{A1}) contains the dynamics
$\dot{\hat{\eta}}=A\hat{\eta}+L_{12}e+N\xi.$ So, the condition ii) implies
that the resulting $\hat{\eta}$ satisfies the unstable matrix differential
equation (\ref{unequation}) asymptotically. Therefore, we achieve the proposed objective.

It is easy to satisfy the condition i) by choosing appropriate gains
$L_{11},L_{12},L_{21},L_{22},L_{3}.$ On the other hand, to satisfy the
condition ii), we introduce the dynamics $\dot{v}=Sv+L_{11}e$ into
(\ref{Augment}), where the matrix $S$ is the same as that in (\ref{exosystem}%
). The idea is inspired by a new viewpoint on the internal model principle
proposed in \cite{Quan(2010)}: $e$ will vanish if it becomes an input of the
internal model such as $\dot{v}=Sv+L_{11}e$, which is further incorporated
into a stable closed-loop linear system. These results are stated in
\textit{Theorems 1-4}.

\textit{Theorem 1}. For (\ref{Augment}), suppose i) $\xi\in\mathcal{L}%
_{\infty}\left(  \left[  0,\infty\right)  ,%
%TCIMACRO{\U{211d} }%
%BeginExpansion
\mathbb{R}
%EndExpansion
\right)  $ is generated by (\ref{exosystem}); ii) the gains $L_{11}%
,L_{12},L_{21},L_{22},L_{3}$ satisfy $\max\operatorname{Re}\lambda\left(
A_{cl}\right)  <0.$ Then $e\rightarrow0$ as $t\rightarrow\infty,$ meanwhile
keeping $x$ bounded. Furthermore, $y=\dot{\hat{\eta}}-A\hat{\eta}%
-N\xi\rightarrow0$ as $t\rightarrow\infty.$

\textit{Proof}\emph{.}\textit{ See Appendix B.}

The key condition of \textit{Theorem 1 }is\textit{ }to find the gains
$L_{11},L_{12},L_{21},L_{22},L_{3}$ satisfying $\max\operatorname{Re}%
\lambda\left(  A_{cl}\right)  <0.$ However, a question immediately arises as
to under what conditions such gains exist for given $S$ and $A$. In
\textit{Theorem 2}, we will answer this question. Denote%
\[
A_{S}=\left[
\begin{array}
[c]{cc}%
S & 0_{m\times n}\\
0_{n\times m} & A
\end{array}
\right]  \in%
%TCIMACRO{\U{211d} }%
%BeginExpansion
\mathbb{R}
%EndExpansion
^{\left(  n+m\right)  \times\left(  n+m\right)  },L_{1}=\left[
\begin{array}
[c]{c}%
L_{11}\\
L_{12}%
\end{array}
\right]  \in%
%TCIMACRO{\U{211d} }%
%BeginExpansion
\mathbb{R}
%EndExpansion
^{n+m}.
\]

\textit{Theorem 2}. If and only if rank$\left(  A_{S}-\lambda I_{n+m}\right)
=n+m-1$ for every eigenvalue $\lambda\in%
%TCIMACRO{\U{2102} }%
%BeginExpansion
\mathbb{C}
%EndExpansion
$ of $A_{S},$ then there exists a vector $L_{1}\in%
%TCIMACRO{\U{211d} }%
%BeginExpansion
\mathbb{R}
%EndExpansion
^{m+n}$ such that the pair $\left(  A_{S},L_{1}\right)  $ is controllable.
Furthermore, if matrix $S$ and $A$ have an eigenvalue in common, then the pair
$\left(  A_{S},L_{1}\right)  $ is uncontrollable for any $L_{1}\in%
%TCIMACRO{\U{211d} }%
%BeginExpansion
\mathbb{R}
%EndExpansion
^{m+n}$.

\textit{Proof}. The first part of \textit{Theorem 2 }can be claimed by
\textit{Lemma 1 }obviously. If matrix $S$ and $A$ have an eigenvalue in
common, denoted by $\lambda_{c}$, then
\begin{align*}
\text{rank}\left(  A_{S}-\lambda_{c}I_{n+m}\right)   &  =\text{rank}\left(
S-\lambda_{c}I_{n}\right)  +\text{rank}\left(  A-\lambda_{c}I_{m}\right) \\
&  \leq m+n-2.
\end{align*}
We can conclude this proof for the second part of \textit{Theorem 2 }by
\textit{Lemma 1. }$\square$

With \textit{Theorems 1-2} in hand, we have

\textit{Theorem 3}. For (\ref{Augment}), suppose i) $\xi\in\mathcal{L}%
_{\infty}\left(  \left[  0,\infty\right)  ,%
%TCIMACRO{\U{211d} }%
%BeginExpansion
\mathbb{R}
%EndExpansion
\right)  $ is generated by (\ref{exosystem})$\ $with appropriate initial
values; ii) rank$\left(  A_{S}-\lambda I\right)  =m+n-1$ for every eigenvalue
$\lambda$ of $A_{S}.$ Then i) there must exist gains $L_{11},L_{12}%
,L_{21},L_{22},L_{3}$ satisfying $\max\operatorname{Re}\lambda\left(
A_{cl}\right)  <0;$ furthermore ii) $e\rightarrow0$ as $t\rightarrow\infty,$
meanwhile keeping $x\left(  t\right)  $ bounded. Moreover, $y=\dot{\hat{\eta}%
}-A\hat{\eta}-N\xi\rightarrow0$ as $t\rightarrow\infty.$

\textit{Proof}. \textit{See Appendix C.}

\textit{Remark 1. }The IID can be given exactly \cite{Francis(1976)}:
$\eta=\Pi w$, where $\Pi\in%
%TCIMACRO{\U{211d} }%
%BeginExpansion
\mathbb{R}
%EndExpansion
^{n\times m}$ satisfies the Sylvester equation $\Pi S=A\Pi+NE^{T}.$ Such
equation has a unique solution if and only if $S$ and $A$ have no eigenvalues
in common \cite[Theorem 13.18, p. 145]{Laub(2005)}. It is easy to see that the
following two conditions are equivalent:%
\[
S\text{ and }A\text{ have no eigenvalues in common}\Leftrightarrow
\text{rank}\left(  A_{S}-\lambda I_{n}\right)  =n+m-1.
\]
Therefore, the solvability condition of the proposed generator is consistent
with that of\ the Sylvester equation $\Pi S=A\Pi+NE^{T}.$ If $A$ is singular,
then $S$ cannot be singular to ensure the existence of the vector $L_{1}\in%
%TCIMACRO{\U{211d} }%
%BeginExpansion
\mathbb{R}
%EndExpansion
^{n+m}$. Unlike the IID generators given by \cite{Shkolnikov(2002)}%
,\cite{Shtessel(2010)}, the proposed IID generator allows $A$ to be singular
in some cases. For example, the pair $\left(  A_{S},L_{1}\right)  $ is
controllable with $A=0,S=\left[
\begin{array}
[c]{cc}%
0 & 1\\
-1 & 0
\end{array}
\right]  ,L_{1}=\left[
\begin{array}
[c]{ccc}%
1 & 1 & 1
\end{array}
\right]  ^{T}.$ This feature broadens the application of the proposed IID generator.

Let us consider that $\xi$ is a vector rather than a scalar, namely%
\begin{equation}
\dot{\eta}=A\eta+\overset{l}{\underset{k=1}{%
%TCIMACRO{\dsum }%
%BeginExpansion
{\displaystyle\sum}
%EndExpansion
}}N_{k}\xi_{k},\eta\left(  0\right)  =0 \label{unequation_multi}%
\end{equation}
where $\eta\in%
%TCIMACRO{\U{211d} }%
%BeginExpansion
\mathbb{R}
%EndExpansion
^{n},\xi_{k}\in%
%TCIMACRO{\U{211d} }%
%BeginExpansion
\mathbb{R}
%EndExpansion
,$ $N_{k}\in%
%TCIMACRO{\U{211d} }%
%BeginExpansion
\mathbb{R}
%EndExpansion
^{n},k=1,\cdots,l$. We have the following result:

\textit{Theorem 4}. For (\ref{unequation_multi}), suppose i) $\xi_{k}%
\in\mathcal{L}_{\infty}\left(  \left[  0,\infty\right)  ,%
%TCIMACRO{\U{211d} }%
%BeginExpansion
\mathbb{R}
%EndExpansion
\right)  $ and can be generated by (\ref{exosystem})$\ $with an appropriate
initial value, $k=1,\cdots,l$; ii) rank$\left(  A_{S}-\lambda I\right)
=m+n-1$ for every eigenvalue $\lambda$ of $A_{S}.$ Then i) there must exist
gains $L_{11},L_{12},L_{21},L_{22},L_{3}$ satisfying $\max\operatorname{Re}%
\lambda\left(  A_{cl}\right)  <0;$ ii) furthermore, the following IID
generator%
\begin{align}
\dot{x}  &  =A_{cl}x+\overset{l}{\underset{k=1}{%
%TCIMACRO{\dsum }%
%BeginExpansion
{\displaystyle\sum}
%EndExpansion
}}N_{cl,k}\xi_{k},x\left(  0\right)  =0\nonumber\\
\hat{\eta}  &  =C_{cl}^{T}x \label{Augment_muti}%
\end{align}
can drive $y=\dot{\hat{\eta}}-A\hat{\eta}-\overset{l}{\underset{k=1}{%
%TCIMACRO{\dsum }%
%BeginExpansion
{\displaystyle\sum}
%EndExpansion
}}N_{k}\xi_{k}\rightarrow0$ as $t\rightarrow\infty,$ meanwhile keeping
$x\left(  t\right)  $ bounded, where $x\in%
%TCIMACRO{\U{211d} }%
%BeginExpansion
\mathbb{R}
%EndExpansion
^{m+n+1},\hat{\eta}\in%
%TCIMACRO{\U{211d} }%
%BeginExpansion
\mathbb{R}
%EndExpansion
^{n},$ $N_{cl,k}=\left[
\begin{array}
[c]{ccc}%
0_{1\times m} & N_{k}^{T} & 0
\end{array}
\right]  ^{T}\in%
%TCIMACRO{\U{211d} }%
%BeginExpansion
\mathbb{R}
%EndExpansion
^{m+n+1},$ $A_{cl},C_{cl}$ are same as in (\ref{Augment}).

\textit{Proof}. By the superposition principle or additive decomposition
\cite{Quan(2009)}, the IID generator (\ref{Augment_muti}) can be decomposed
into%
\begin{align}
\dot{x}_{k}  &  =A_{cl}x_{k}+N_{cl,k}\xi_{k},x_{k}\left(  0\right)
=0\nonumber\\
\hat{\eta}_{k}  &  =C_{cl}^{T}x_{k},k=1,\cdots,l \label{decomposedp}%
\end{align}
with the relation%
\begin{equation}
x=\overset{l}{\underset{k=1}{%
%TCIMACRO{\dsum }%
%BeginExpansion
{\displaystyle\sum}
%EndExpansion
}}x_{k},\hat{\eta}=\overset{l}{\underset{k=1}{%
%TCIMACRO{\dsum }%
%BeginExpansion
{\displaystyle\sum}
%EndExpansion
}}\hat{\eta}_{k}. \label{relationship}%
\end{equation}
By conditions i)-ii) and \textit{Theorem 3},\textit{ }the IID generator
(\ref{decomposedp}) for each $\xi_{k}\left(  t\right)  $ can drive $y_{k}%
=\dot{\hat{\eta}}_{k}-A\hat{\eta}_{k}-N_{k}\xi_{k}\rightarrow0$ as
$t\rightarrow\infty,$ meanwhile keeping $x_{k}\left(  t\right)  $ bounded. By
(\ref{relationship}), we have
\[
y=\dot{\hat{\eta}}-A\hat{\eta}-\overset{l}{\underset{k=1}{%
%TCIMACRO{\dsum }%
%BeginExpansion
{\displaystyle\sum}
%EndExpansion
}}N_{k}\xi_{k}=\overset{l}{\underset{k=1}{%
%TCIMACRO{\dsum }%
%BeginExpansion
{\displaystyle\sum}
%EndExpansion
}}\left(  \dot{\hat{\eta}}_{k}-A\hat{\eta}_{k}-N_{k}\xi_{k}\right)
\rightarrow0
\]
as $t\rightarrow\infty,$ meanwhile keeping $x\left(  t\right)  =%
%TCIMACRO{\dsum \nolimits_{k=1}^{l}}%
%BeginExpansion
{\displaystyle\sum\nolimits_{k=1}^{l}}
%EndExpansion
x_{k}\left(  t\right)  $ bounded. $\square$

\section{$\mathcal{H}_{2}/\mathcal{H}_{\infty}$ Optimal Design of IID
Generator}

So far, we have proposed the structure of the IID generators, and further
investigated the existence of their parameters $L_{11},L_{12},L_{21}%
,L_{22},L_{3}.$ However, there exist infinite choices of the parameters
$L_{11},L_{12},L_{21},L_{22},L_{3}$ to satisfy$\ \max\operatorname{Re}%
\lambda\left(  A_{cl}\right)  <0.$ In this section, we will design these
parameters according to some optimization principles.

In practice, the forcing term $\xi$ often cannot be modeled as
(\ref{exosystem}) without perturbation. Assume $\varepsilon\in%
%TCIMACRO{\U{211d} }%
%BeginExpansion
\mathbb{R}
%EndExpansion
$ to be a bounded perturbation. Driven by $\xi+\varepsilon,$ the solution to
(\ref{Augment}) satisfies%
\begin{align}
\dot{x}_{\varepsilon}  &  =A_{cl}x_{\varepsilon}+N_{cl}\left(  \xi
+\varepsilon\right)  ,x_{\varepsilon}\left(  0\right)  =0\nonumber\\
\hat{\eta}_{\varepsilon}  &  =C_{cl}^{T}x_{\varepsilon}. \label{solution}%
\end{align}
We expect to design the parameters $L_{11},L_{12},L_{21},L_{22},L_{3}$ such
that $\hat{\eta}_{\varepsilon}-\hat{\eta}$ is not sensitive to the
perturbation $\varepsilon.$ Subtracting (\ref{Augment}) from (\ref{solution})
results in%
\begin{align}
\dot{x}_{e}  &  =A_{cl}x_{e}+N_{cl}\varepsilon,x_{e}\left(  0\right)
=0\nonumber\\
\hat{\eta}_{e}  &  =C_{cl}^{T}x_{e} \label{errors}%
\end{align}
where $\hat{\eta}_{e}=\hat{\eta}_{\varepsilon}-\hat{\eta}\ $and$\ x_{e}%
=x_{\varepsilon}-x.$ Denote%
\[
A_{S}^{\prime}=\left[
\begin{array}
[c]{cc}%
A_{S} & L_{1}\\
0_{1\times\left(  m+n\right)  } & 0
\end{array}
\right]  ,L_{23}=\left[
\begin{array}
[c]{ccc}%
L_{21} & L_{22} & L_{3}%
\end{array}
\right]  ^{T}.
\]
Then $A_{cl}=A_{S}^{\prime}+B_{1}L_{23}^{T}.$ The system (\ref{errors}) can be
rewritten as%
\begin{align*}
\dot{x}_{e}  &  =A_{S}^{\prime}x_{e}+B_{1}u+N_{cl}\varepsilon,x_{e}\left(
0\right)  =0\\
\hat{\eta}_{e}  &  =C_{cl}^{T}x_{e}\\
u  &  =L_{23}^{T}x_{e}%
\end{align*}
which is shown in Fig.1.\begin{figure}[h]
\begin{center}
\includegraphics[
scale=1.2 ]{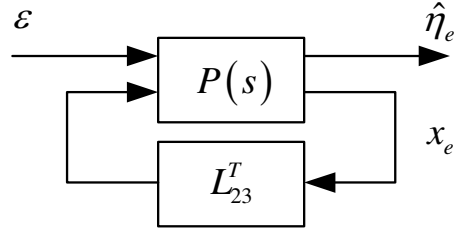}
\end{center}
\caption{State-feedback control}%
\end{figure}

Although some tracking and robustness are best captured by an $\mathcal{H}%
_{\infty}$ criterion, noise insensitivity is more naturally expressed by the
$\mathcal{H}_{2}$ criterion. Robust pole placement specifications are also
required for\ reasonable feedback gains. Denote by $T_{\hat{\eta}%
_{e}\varepsilon}$ the closed-loop transfer functions from $\varepsilon$ to
$\hat{\eta}_{e}.$ For simplicity, we determine $L_{1}$ similar to (\ref{L1})
beforehand. Then, our goal is to design a state-feedback law $u=L_{23}%
^{T}x_{e}$ that

\textbullet\ Maintains $\left\Vert T_{\hat{\eta}_{e}\varepsilon}\right\Vert
_{\infty}$ below some prescribed value $\gamma_{0}>0.$

\textbullet\ Maintains $\left\Vert T_{\hat{\eta}_{e}\varepsilon}\right\Vert
_{2}$ below some prescribed value $\nu_{0}>0.$

\textbullet\ Minimizes an $\mathcal{H}_{2}/\mathcal{H}_{\infty}$ trade-off
criterion of the form $\alpha\left\Vert T_{\hat{\eta}_{e}\varepsilon
}\right\Vert _{\infty}+\beta\left\Vert T_{\hat{\eta}_{e}\varepsilon
}\right\Vert _{2},\alpha\geq0,\beta\geq0.$

\textbullet\ Places the closed-loop poles in a prescribed region $\mathcal{D}$
of the open left-half plane.

Formally, the objective is to find $L_{23}$ such that:%
\begin{equation}%
\begin{array}
[c]{ll}%
\underset{L_{23}}{\min} & \alpha\left\Vert T_{\hat{\eta}_{e}\varepsilon
}\right\Vert _{\infty}+\beta\left\Vert T_{\hat{\eta}_{e}\varepsilon
}\right\Vert _{2}\\
\text{s.t.} & \left\Vert T_{\hat{\eta}_{e}\varepsilon}\right\Vert _{\infty
}<\gamma_{0}\\
& \left\Vert T_{\hat{\eta}_{e}\varepsilon}\right\Vert _{2}<\nu_{0}\\
& \lambda\left(  A_{cl}\right)  \in\mathcal{D}=\left\{  z\left.  \in%
%TCIMACRO{\U{2102} }%
%BeginExpansion
\mathbb{C}
%EndExpansion
\right\vert Q+Mz+M\bar{z}<0\right\}
\end{array}
\label{optim}%
\end{equation}
where matrices $Q=Q^{T}$ and $M$ is a suitable matrix.

\textit{Remark 2.} The perturbation is not necessary to be $\varepsilon
\in\mathcal{L}_{2}\ $in practice although $\mathcal{H}_{2}$ optimization is
considered. From (\ref{Augment}), the state is still bounded if $\varepsilon
\ $is bounded and $\max\operatorname{Re}\lambda\left(  A_{cl}\right)  <0.$

\textit{Remark 3}. The MATLAB function \textquotedblleft
msfsyn\textquotedblright\cite{Gahinet(1995)}\ is applicable to solve the
optimization problem (\ref{optim}).

\section{Simulation Examples}

For simplicity, in the following examples, the prescribed region $\mathcal{D}$
of the open left-half plane is chosen to be an intersection of a conic sector
centered at the origin with inner angle $\frac{3\pi}{4}$ and a vertical strip
$\left[  -10,-1\right]  ,$ shown in Fig.2.\begin{figure}[h]
\begin{center}
\includegraphics[
scale= 0.8]{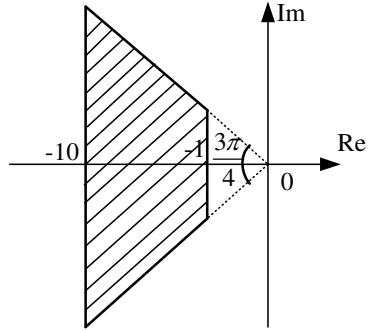}
\end{center}
\caption{Prescribed region $\mathcal{D}$}%
\end{figure}

\textit{Example 1}. In (\ref{unequation}), $A=0,N=1,$ where $\xi$ is generated
by (\ref{exosystem}) with $S=\left[
\begin{array}
[c]{cc}%
0 & 1\\
-1 & 0
\end{array}
\right]  ,E=\left[
\begin{array}
[c]{c}%
1\\
0
\end{array}
\right]  ,w\left(  0\right)  =\left[
\begin{array}
[c]{c}%
1\\
1
\end{array}
\right]  .$

Since $A_{S}$ has three different eigenvalues $0,\pm j,$ rank$\left(
A_{S}-\lambda I\right)  =2$ for $\lambda=0,\pm j.$ Obviously, the dynamic IID
generators proposed in \cite{Shkolnikov(2002)} and \cite{Shtessel(2010)} are
inapplicable to this example. Similar to (\ref{L1}) in \emph{Appendix A},
$L_{1}$ is chosen as $L_{1}=\left[
\begin{array}
[c]{ccc}%
1 & 0 & 1
\end{array}
\right]  ^{T}$. Choosing $\gamma_{0}=20,\nu_{0}=20,\alpha=0.5,\beta=0.5\ $and
solving (\ref{optim}) by the MATLAB function \textquotedblleft
msfsyn\textquotedblright, we obtain $L_{23}=10^{3}\times\left[
\begin{array}
[c]{cccc}%
0.5360 & 1.0746 & -0.9743 & -0.0219
\end{array}
\right]  ^{T}\ $with $\left\Vert T_{\hat{\eta}_{e}\varepsilon}\right\Vert
_{\infty}=1.75$ and $\left\Vert T_{\hat{\eta}_{e}\varepsilon}\right\Vert
_{2}=2.61.$ By solving (\ref{Augment}) in forward time, the IID is obtained.
As shown in Fig.3, the one-dimensional estimated IID $\hat{\eta}$ is bounded
and $y=\dot{\hat{\eta}}-A\hat{\eta}-N\xi\rightarrow0$ as $t\rightarrow\infty.$
Moreover, it is easy to see that the estimated IID $\hat{\eta}$ converges to
the desired IID. In the presence of $\varepsilon,$ as shown in Fig.4, it is
easy to see that the estimated IID can also converge to the desired IID with a
small error.\begin{figure}[h]
\begin{center}
\includegraphics[
scale=0.65]{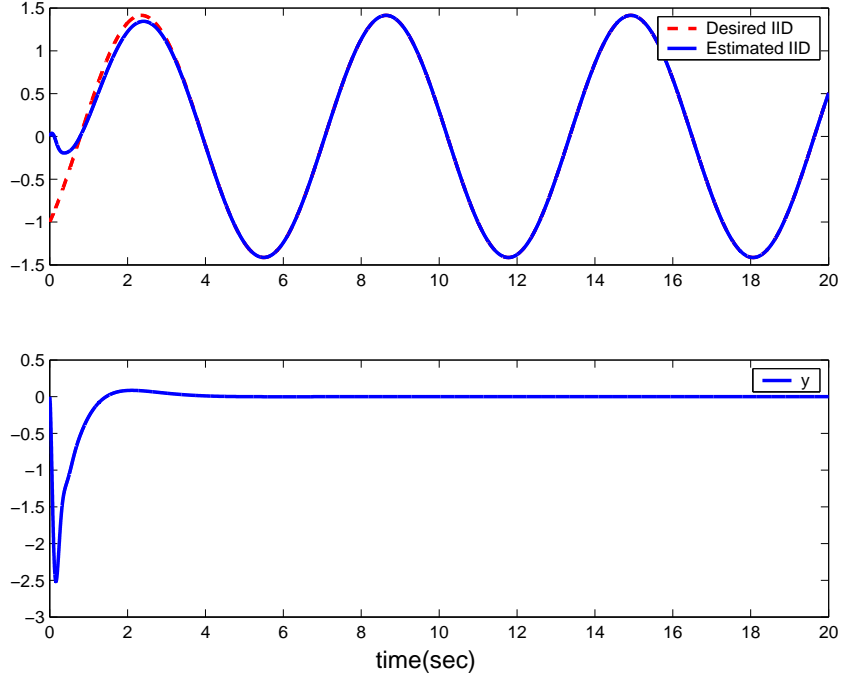}
\end{center}
\caption{Output of the IID generator in \textit{Example 1}}%
\end{figure}\begin{figure}[h]
\begin{center}
\includegraphics[
scale=0.65]{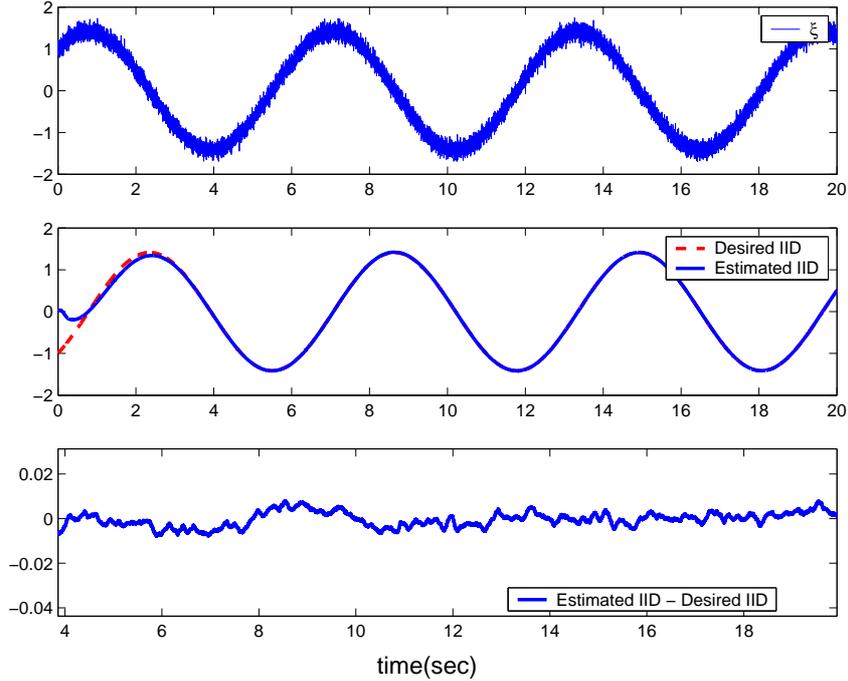}
\end{center}
\caption{Output of the IID generator in \textit{Example 1 }in the presence of
noise}%
\end{figure}

\textit{Example 2}. In (\ref{unequation_multi}),
\begin{equation}
A\left(  t\right)  =\left[
\begin{array}
[c]{cc}%
0.2\sin(0.05t) & 1\\
-1 & 1
\end{array}
\right]  ,N_{1}=\left[
\begin{array}
[c]{c}%
1\\
2
\end{array}
\right]  ,N_{2}=\left[
\begin{array}
[c]{c}%
0\\
1
\end{array}
\right]  ,l=2 \label{Example 2}%
\end{equation}
where $\xi_{1},\xi_{2}$ are generated respectively by (\ref{exosystem}) with
\[
S=\left[
\begin{array}
[c]{cc}%
0 & 0.2\\
-0.2 & 0
\end{array}
\right]  ,E=\left[
\begin{array}
[c]{c}%
1\\
0
\end{array}
\right]  ,w_{1}\left(  0\right)  =\left[
\begin{array}
[c]{cc}%
1 & 0
\end{array}
\right]  ^{T},w_{2}\left(  0\right)  =\left[
\begin{array}
[c]{cc}%
0 & 1
\end{array}
\right]  ^{T}.
\]

In the IID generator (\ref{Augment_muti}), $A$ will be replaced by $A\left(
t\right)  $ in (\ref{Example 2}) to obtain an approximate IID. However, for
sake of designing $L_{1}$ and $L_{23}$, we consider $A\left(  t\right)
\equiv\left[
\begin{array}
[c]{cc}%
0 & 1\\
-1 & 1
\end{array}
\right]  $ first. Similar to (\ref{L1}) in \emph{Appendix A}, $L_{1}$ is
designed as $L_{1}=\left[
\begin{array}
[c]{cccc}%
1 & 0 & 1 & 0.0670
\end{array}
\right]  ^{T}$. Choosing $\gamma_{0}=20,\nu_{0}=20,\alpha=0.5,\beta=0.5\ $and
solving (\ref{optim}) by the MATLAB function \textquotedblleft
msfsyn\textquotedblright, we obtain $L_{23}=$ $10^{4}\times\left[
\begin{array}
[c]{ccccc}%
-0.5702 & 1.0009 & 0.5159 & 0.0850 & -0.0025
\end{array}
\right]  ^{T}\ $with $\left\Vert T_{\hat{\eta}_{e}\varepsilon}\right\Vert
_{\infty}=9.37$ and $\left\Vert T_{\hat{\eta}_{e}\varepsilon}\right\Vert
_{2}=16.$ By solving the resulting IID generator (\ref{Augment_muti}) in
forward time, the estimated IID is obtained. As shown in Fig.5, the
two-dimensional estimated IID $\hat{\eta}$ is bounded, and each element of
$y=\dot{\hat{\eta}}-A\hat{\eta}-N\xi\in%
%TCIMACRO{\U{211d} }%
%BeginExpansion
\mathbb{R}
%EndExpansion
^{2}\ $is bounded ultimately by a very small positive value.

\begin{figure}[h]
\begin{center}
\includegraphics[
scale=0.65]{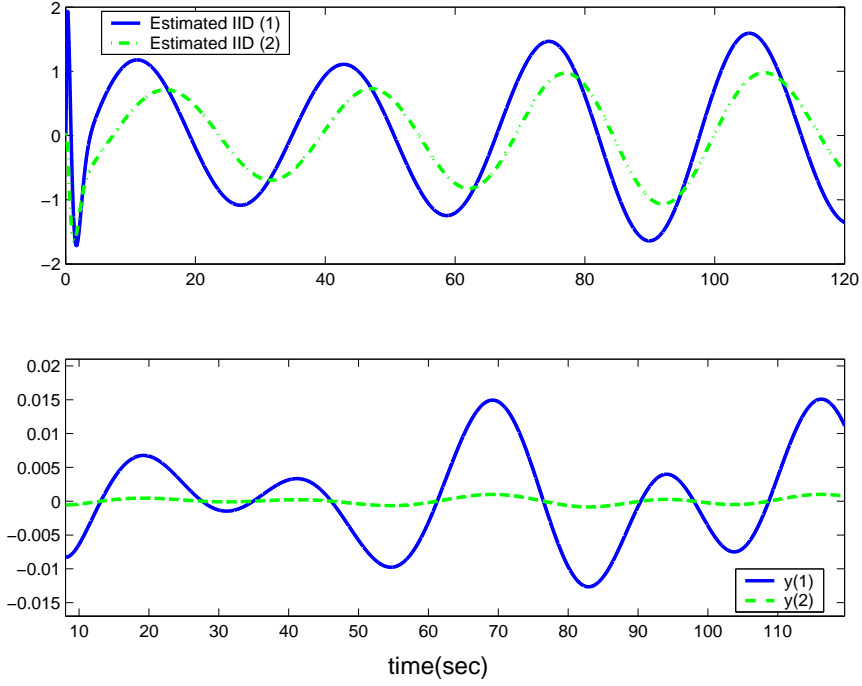}
\end{center}
\caption{Output of the IID generator in \textit{Example 2 }with a time-varying
matrix}%
\end{figure}

\section{Conclusions}

In this paper, a new causal dynamic IID generator is proposed. By solving it
in forward time, the IID can be obtained. Owing to the dynamics, it can
suppress noise and perturbations. Compared with the existing similar
generators, it is applicable to the singular case and can easily be extended
to slowly time-varying unstable matrix differential equations in the same
framework without extra computation. The simulation examples demonstrate the
effectiveness of the proposed IID generator.

\section{Appendix}

\subsection{Proof of Lemma 1}

Before presenting the proof, we introduce a lemma.

\textit{Lemma 2 (PBH controllability test)} \cite[Theorem 4.8, p.102]%
{Terrell(2009)}. The matrix pair $\left(  F,B\right)  $ is controllable if and
only if%
\[
\text{rank}\left[
\begin{array}
[c]{cc}%
F-\lambda I & B
\end{array}
\right]  =n
\]
for every eigenvalue $\lambda\in%
%TCIMACRO{\U{2102} }%
%BeginExpansion
\mathbb{C}
%EndExpansion
$ of $F$.

\textit{Sufficiency of Lemma 1 (A Constructive Proof). }For\textit{ }$F\in%
%TCIMACRO{\U{211d} }%
%BeginExpansion
\mathbb{R}
%EndExpansion
^{n\times n},$ there exists a matrix $T\in%
%TCIMACRO{\U{211d} }%
%BeginExpansion
\mathbb{R}
%EndExpansion
^{n\times n}$ such that \cite[Theorem 9.22, pp.82-83]{Laub(2005)}%
\[
T^{-1}FT=J=\text{diag}\left(  J_{1},\cdots,J_{n_{s}}\right)
\]
where each of the Jordan block matrices $J_{1},\cdots,J_{n_{s}}$ is of the
form%
\begin{equation}
J_{i}=\left[
\begin{array}
[c]{cccc}%
\lambda_{i} & 1 &  & \\
& \lambda_{i} & \ddots & \\
&  & \ddots & 1\\
&  &  & \lambda_{i}%
\end{array}
\right]  \label{J1}%
\end{equation}
in the case of real eigenvalues $\lambda_{i},$ and%
\begin{equation}
J_{i}=\left[
\begin{array}
[c]{cccc}%
M_{i} & I_{2} &  & \\
& M_{i} & \ddots & \\
&  & \ddots & I_{2}\\
&  &  & M_{i}%
\end{array}
\right]  \label{J2}%
\end{equation}
where $M_{i}=\left[
\begin{array}
[c]{cc}%
\alpha_{i} & \beta_{i}\\
-\beta_{i} & \alpha_{i}%
\end{array}
\right]  $ and $I_{2}=\left[
\begin{array}
[c]{cc}%
1 & 0\\
0 & 1
\end{array}
\right]  $ in the case of $\alpha_{i}\pm j\beta_{i},\beta_{i}\neq0.$ For
simplicity and without loss of generality, we assume that only the last Jordan
block $J_{n_{s}}$\ is in the form of (\ref{J2}). The Jordan block $J_{i}%
$\ corresponds to a left eigenvector $v_{i}\in%
%TCIMACRO{\U{211d} }%
%BeginExpansion
\mathbb{R}
%EndExpansion
^{n},i=1,\cdots,n_{s}-1$, and $J_{n_{s}}$ corresponds to a couple of left
eigenvectors $v_{n_{s}}\in%
%TCIMACRO{\U{2102} }%
%BeginExpansion
\mathbb{C}
%EndExpansion
^{n},v_{n_{s}+1}\in%
%TCIMACRO{\U{2102} }%
%BeginExpansion
\mathbb{C}
%EndExpansion
^{n}$. It is easy to see that $v_{i}^{H}v_{k}=0$ by the form of $J$, $i\neq
k$, except for $v_{n_{s}}$ and $v_{n_{s}+1}.$ Every eigenvalue $\lambda_{i}\in%
%TCIMACRO{\U{2102} }%
%BeginExpansion
\mathbb{C}
%EndExpansion
$ corresponds to a left eigenvector $0\neq v_{i}\in%
%TCIMACRO{\U{2102} }%
%BeginExpansion
\mathbb{C}
%EndExpansion
^{n}$ such that $v_{i}^{H}J=\lambda_{i}v_{i}^{H},i=1,\cdots,n_{s}+1,$ which
implies that$\ \bar{v}^{H}J=\bar{\lambda}\bar{v}^{H}.$ Here $\bar{x}$
represents the element-by-element conjugation of $x\in%
%TCIMACRO{\U{2102} }%
%BeginExpansion
\mathbb{C}
%EndExpansion
^{n},$ and $x^{H}$ represents the conjugate transpose of $x\in%
%TCIMACRO{\U{2102} }%
%BeginExpansion
\mathbb{C}
%EndExpansion
^{n}.$ Therefore, for a couple of conjugate complex roots, their eigenvectors
can be chosen to be conjugate, namely $v_{n_{s}+1}=\bar{v}_{n_{s}},$ so that%
\begin{equation}
B=T\sum_{i=1}^{n_{s}+1}v_{i} \label{L1}%
\end{equation}
is a real vector. Next, we will show that rank$\left[
\begin{array}
[c]{cc}%
F-\lambda I_{n} & B
\end{array}
\right]  =n\ $for every eigenvalue $\lambda\in%
%TCIMACRO{\U{2102} }%
%BeginExpansion
\mathbb{C}
%EndExpansion
$ of $F$. Suppose, to the contrary, that there exists a vector $0\neq p\in%
%TCIMACRO{\U{2102} }%
%BeginExpansion
\mathbb{C}
%EndExpansion
^{n}$ and $\lambda_{k}\in%
%TCIMACRO{\U{2102} }%
%BeginExpansion
\mathbb{C}
%EndExpansion
$ such that $p^{H}\left[
\begin{array}
[c]{cc}%
F-\lambda_{k}I_{n} & B
\end{array}
\right]  =0,$ namely%
\begin{align*}
p^{H}\left(  F-\lambda_{k}I_{n}\right)   &  =0\\
p^{H}B  &  =0.
\end{align*}
Furthermore, we have%
\begin{align}
p^{H}T\left(  J-\lambda_{k}I_{n}\right)   &  =0\label{e1}\\
p^{H}T\sum_{i=1}^{n_{s}+1}v_{i}  &  =0. \label{e2}%
\end{align}
Since rank$\left(  F-\lambda I_{n}\right)  =$rank$\left(  J-\lambda
I_{n}\right)  =n-1$ for every eigenvalue of $F,$ each eigenvalue corresponds
to exactly one eigenvector. As a result, the equation (\ref{e1}) implies
$T^{H}p=\mu v_{k},0\neq\mu\in%
%TCIMACRO{\U{2102} }%
%BeginExpansion
\mathbb{C}
%EndExpansion
.$ Furthermore, the equation (\ref{e2}) implies%
\begin{equation}
\mu v_{k}^{H}v_{k}=0,k=1,\cdots,n_{s}-1 \label{r1}%
\end{equation}
or
\begin{equation}
\mu v_{n_{s}}^{H}\left(  v_{n_{s}}+\bar{v}_{n_{s}}\right)  =0,k=n_{s}
\label{r2}%
\end{equation}
or%
\begin{equation}
\mu\bar{v}_{n_{s}}^{H}\left(  v_{n_{s}}+\bar{v}_{n_{s}}\right)  =0,k=n_{s}+1
\label{r3}%
\end{equation}
where the orthogonality and $v_{n_{s}+1}=\bar{v}_{n_{s}}$ have been utilized.
The equation (\ref{r1}) implies that%
\[
v_{k}=0
\]
which contradicts with $v_{i}\neq0_{n},i=1,\cdots,n_{s}+1.$ The equation
(\ref{r2}) or (\ref{r3}) implies that
\[
v_{n_{s}}+\bar{v}_{n_{s}}=0.
\]
Consequently, $v_{n_{s}}\ $is in the form of $v_{n_{s}}=jv_{n_{s}}^{r},$ where
$v_{n_{s}}^{r}\in%
%TCIMACRO{\U{211d} }%
%BeginExpansion
\mathbb{R}
%EndExpansion
^{n}.$ Since
\begin{align*}
v_{n_{s}}^{H}J  &  =\left(  \alpha_{n_{s}}+j\beta_{n_{s}}\right)  v_{n_{s}%
}^{H}\\
\bar{v}_{n_{s}}^{H}J  &  =\left(  \alpha_{n_{s}}-j\beta_{n_{s}}\right)
\bar{v}_{n_{s}}^{H}%
\end{align*}
we have%
\begin{align*}
0  &  =\left(  v_{n_{s}}^{H}+\bar{v}_{n_{s}}^{H}\right)  J\\
&  =\left(  \alpha_{n_{s}}+j\beta_{n_{s}}\right)  v_{n_{s}}^{H}+\left(
\alpha_{n_{s}}-j\beta_{n_{s}}\right)  \bar{v}_{n_{s}}^{H}\\
&  =2\beta_{n_{s}}v_{n_{s}}^{r}j
\end{align*}
which contradicts with $\beta_{n_{s}}\neq0\ $or$\ v_{n_{s}}\neq0_{n}%
.$\ Therefore, rank$\left[
\begin{array}
[c]{cc}%
F-\lambda I_{n} & B
\end{array}
\right]  =n\ $for every eigenvalue $\lambda\in%
%TCIMACRO{\U{2102} }%
%BeginExpansion
\mathbb{C}
%EndExpansion
$ of $F$, namely the pair $\left(  F,B\right)  $ is controllable by
\textit{Lemma 2}.

\textit{Necessity of Lemma 1}. If rank$\left(  F-\lambda I_{n}\right)  \neq
n-1,$ namely rank$\left(  F-\lambda I_{n}\right)  \leq n-2$ for every
eigenvalue $\lambda$ of $F,$ then rank$\left[
\begin{array}
[c]{cc}%
F-\lambda I_{n} & B
\end{array}
\right]  \leq n-1$ for any $B\in%
%TCIMACRO{\U{211d} }%
%BeginExpansion
\mathbb{R}
%EndExpansion
^{n},$ namely the pair $\left(  F,B\right)  $ is uncontrollable by
\textit{Lemma 2}.

\subsection{Proof of Theorem 1}

Before proving, we introduce a lemma.

\textit{Lemma 3}. If the pair $\left(  F,B\right)  $ is controllable, then
there exists a vector $C\in%
%TCIMACRO{\U{211d} }%
%BeginExpansion
\mathbb{R}
%EndExpansion
^{n}$ such that%
\[
C^{T}(sI_{n}-F)^{-1}B=\frac{1}{\det\left(  sI_{n}-F\right)  }%
\]
where $F\in%
%TCIMACRO{\U{211d} }%
%BeginExpansion
\mathbb{R}
%EndExpansion
^{n\times n}\ $and$\ B\in%
%TCIMACRO{\U{211d} }%
%BeginExpansion
\mathbb{R}
%EndExpansion
^{n}.$

\textit{Proof.} First, we have $(sI_{n}-F)^{-1}B=G\left[
\begin{array}
[c]{ccc}%
s^{n-1} & \cdots & 1
\end{array}
\right]  ^{T}\left/  \det\left(  sI_{n}-F\right)  \right.  $, where $G\in%
%TCIMACRO{\U{211d} }%
%BeginExpansion
\mathbb{R}
%EndExpansion
^{n\times n}.$ If the pair $\left(  F,B\right)  $ is controllable, then the
matrix $G$ is of full rank \cite{Cao (2008)}. We can complete this proof by
choosing $C=\left(  G^{-1}\right)  ^{T}\left[
\begin{array}
[c]{cccc}%
0 & \cdots & 0 & 1
\end{array}
\right]  ^{T}.$ $\square$

\textit{Proof of Theorem 1}. The IID generator (\ref{Augment}) contains the
dynamics $\dot{v}=Sv+L_{11}e.$ Its Laplace transformation is%
\[
v\left(  s\right)  =\left(  sI_{m}-S\right)  ^{-1}L_{11}e\left(  s\right)  .
\]
The condition $\max\operatorname{Re}\lambda\left(  A_{cl}\right)  <0$ implies
that the pair $\left(  S,L_{11}\right)  $ is controllable. Further by
\textit{Lemma 3}, there exists a vector$\ C_{e}\in%
%TCIMACRO{\U{211d} }%
%BeginExpansion
\mathbb{R}
%EndExpansion
^{m}$ such that%
\[
C_{e}^{T}v\left(  s\right)  =C_{e}^{T}\left(  sI_{m}-S\right)  ^{-1}%
L_{11}e\left(  s\right)  =\frac{1}{\det\left(  sI_{m}-S\right)  }e\left(
s\right)
\]
namely,%
\begin{equation}
e\left(  s\right)  =\det\left(  sI_{m}-S\right)  C_{e}^{T}v\left(  s\right)  .
\label{e}%
\end{equation}
By (\ref{Augment}), the transfer function from $\xi$ to $v\ $is%
\[
v\left(  s\right)  =C_{v}^{T}\left(  sI_{m+n+1}-A_{cl}\right)  ^{-1}N_{cl}%
\xi\left(  s\right)
\]
where $C_{v}=\left[
\begin{array}
[c]{ccc}%
I_{m} & 0_{m\times n} & 0
\end{array}
\right]  ^{T}.$ Substituting the equation above into (\ref{e}) yields
\[
e\left(  s\right)  =\det\left(  sI_{m}-S\right)  C_{e}^{T}C_{v}^{T}\left(
sI_{m+n+1}-A_{cl}\right)  ^{-1}N_{cl}\xi\left(  s\right)  .
\]
Since $\xi$ is generated by (\ref{exosystem}), we have$\ \xi\left(  s\right)
=E^{T}\left(  sI_{m}-S\right)  ^{-1}w\left(  0\right)  ,$ where $w\left(
0\right)  \in%
%TCIMACRO{\U{211d} }%
%BeginExpansion
\mathbb{R}
%EndExpansion
^{m}.$ Since $\left(  sI_{m}-S\right)  ^{-1}=\frac{1}{\det\left(
sI_{m}-S\right)  }$adj$\left(  sI_{m}-S\right)  $, $e\left(  s\right)  $ is
further represented as%
\begin{align}
e\left(  s\right)   &  =\det\left(  sI_{m}-S\right)  C_{e}^{T}C_{v}^{T}\left(
sI_{m+n+1}-A_{cl}\right)  ^{-1}N_{cl}E^{T}\frac{1}{\det\left(  sI_{m}%
-S\right)  }\text{adj}\left(  sI_{m}-S\right)  w\left(  0\right) \nonumber\\
&  =C_{e}^{T}C_{v}^{T}\left(  sI_{m+n+1}-A_{cl}\right)  ^{-1}N_{cl}%
E^{T}\text{adj}\left(  sI_{m}-S\right)  w\left(  0\right)  . \label{error}%
\end{align}
Since $\max\operatorname{Re}\lambda\left(  A_{cl}\right)  <0\ $and the order
of $A_{cl}$ is higher than that of $S,$ for any initial values $w\left(
0\right)  ,$ we have $e\rightarrow0$ as $t\rightarrow\infty$ from
(\ref{error}). Since $\xi$ is bounded on $\left[  0,\infty\right)  $ and
$\max\operatorname{Re}\lambda\left(  A_{cl}\right)  <0$, the signals $v$ and
$\hat{\eta}$ in (\ref{Augment}) are bounded. Since the IID (\ref{Augment})
contains the relation $\dot{\hat{\eta}}=A\hat{\eta}+L_{12}e+N\xi.$ By the
obtained result that $e\rightarrow0$ as $t\rightarrow\infty,$ we have
$y=L_{12}e=\dot{\hat{\eta}}-A\hat{\eta}-N\xi\rightarrow0$ as $t\rightarrow
\infty.$

\subsection{Proof of Theorem 3}

By condition ii) and \textit{Theorem 2}, there exists a vector $L_{1}$ such
that the pair $\left(  A_{S},L_{1}\right)  $ is controllable. Consider the
pair%
\begin{equation}
\left(  \left(
\begin{array}
[c]{cc}%
A_{S} & L_{1}\\
0_{1\times\left(  n+m\right)  } & 0
\end{array}
\right)  ,\left(
\begin{array}
[c]{c}%
0_{n+m}\\
1
\end{array}
\right)  \right)  . \label{pair}%
\end{equation}
The controllability matrix of pair (\ref{pair}) is%
\[
W=\left(
\begin{array}
[c]{ccccc}%
0_{n+m} & L_{1} & A_{S}L_{1} & \cdots & A_{S}^{n+m-1}L_{1}\\
1 & 0 & 0 & \cdots & 0
\end{array}
\right)  .
\]
Since the pair $\left(  A_{S},L_{1}\right)  $ is controllable, rank$\left(
\begin{array}
[c]{ccccc}%
0_{n+m} & L_{1} & A_{S}L_{1} & \cdots & A_{S}^{n+m-1}L_{1}%
\end{array}
\right)  =n+m.$ Consequently, rank$W=n+m+1.$ Therefore, the pair (\ref{pair})
is controllable, namely there must exist gains $L_{21},L_{22},L_{3}$ such that
$\max\operatorname{Re}\lambda\left(  A_{cl}\right)  <0.$ The remainder of
proof is the Due to as \textit{Theorem 1}.

\end{document}